\documentclass{article}
\usepackage[dvips]{epsfig}
\usepackage[dvips]{graphics}
\begin{document}

\title{Superheavy particles either for UHECR or for muon anomaly}
\author{H. Ch\'avez\thanks{CNPq fellow at Centro Brasileiro de Pesquisas F\'{\i}sicas, Rio de
Janeiro, Brazil}, L. Masperi\thanks{On leave of absence from
Centro
At\'omico Bariloche, Argentina},  and M. Orsaria\thanks{CAPES fellow at Centro Brasileiro de Pesquisas F\'{\i}sicas, Rio de Janeiro, Brazil}%
\\ Centro Latinoamericano de F\'{\i}sica,\\Av. Venceslau Br\'az 71
Fundos, 22290-140 Rio de Janeiro, Brazil}%
\date{}%
\maketitle
\begin{abstract}
We show that, according to the scheme of spontaneous breakings
starting from a GUT with symmetry $E_6$, it is possible that
either a superheavy particle without ordinary interactions is
source of ultra-high energy cosmic rays or a not so heavy lepton
mixes with muon explaining the recently observed discrepancy of
the anomalous magnetic moment of the latter.
\end{abstract}
\section{Introduction}

To the already much discussed problem of the roughly isotropic
cosmic rays with energy above $10^{20}$ eV~\cite{Watson}, one may
add the recently observed discrepancy~\cite{Brown} in the muon
anomalous magnetic moment (MAM) as requiring theories beyond the
Standard Model (SM) of elementary particles.

There are many explanations of ultra-high energy cosmic rays
(UHECR) based on decaying superheavy objects with lifetime larger
than the universe age~\cite{Sigl}, as well as possible
contributions of new particles to add to the theoretical
evaluation of MAM and fill the $2.6-\sigma$ gap up to the
experimental value~\cite{Einhorn}. But since the simplest solution
of the latter problem is to include a heavy lepton which mixes
with muon, one may inquire whether the same scheme offers a
superheavy particle which might be origin of UHECR. In the frame
of Grand Unification Theories (GUT) the most convenient symmetry
is $E_6$, which may come down from the more fundamental
superstring theory, because it contains for each generation a
heavy charged lepton and one particle without ordinary
interactions and therefore with possible great stability.

The feasibility of the model depends on details of the Higgs
fields which produce the breaking of symmetry from $E_6$ down to
QCD and electromagnetism giving mass subsequently to superheavy
particles and to ordinary ones.

We will show that there are essentially two interesting
alternative chains. In the first the particle without ordinary
interactions is heavier than the exotic lepton and can decay in it
through non-standard gauge bosons, having therefore a short
lifetime with no possibility of explaining the UHECR. But since
the exotic doublet of leptons mixes strongly with the ordinary
one, the muon acquires a relevant flavour-changing coupling with
Higgs which may give a contribution to the MAM of the order of the
present discrepancy with the experimental value. The second
possible chain gives an exotic lepton with a mass larger than that
of the particle without ordinary couplings and with a weak mixing
with light particles, so that it contributes negligibly to MAM.
But now the particle without ordinary couplings decays through
virtual exotic fermions whose extremely low mixing with light ones
may produce a lifetime as large as the universe age allowing it as
source of UHECR.

\section{$E_6$ and its breaking}

The GUT model based on the symmetry of the exceptional group $E_6$
has 78 gauge bosons of which 45 are those predicted by $SO(10)$
and the rest will be called $X$. The left-handed fermions are
normally placed in the fundamental 27-dimensional representation,
being the ordinary ones including $\nu^c$ in the representation 16
of $SO(10)$, an exotic lepton doublet
$(\begin{array}{c}N\\E\end{array})$ together with the singlets
$N^c$ and $E^c$ and those of charge $-\frac{1}{3}$ quarks $D$ and
$D^c$ in a representation 10, and finally a fermion $L$ without
interactions with the 45 $SO(10)$ bosons in the trivial
representation 1. As it is usual, we work with the charge
conjugated left components instead of the right-handed ones.

$E_6$ symmetry must break at a high scale to that of the SM, and
the latter to the present one of QCD and electromagnetism at low
electroweak (EW) scale. We will assume a detailed high scale chain
passing through the maximum subgroups involving the intermediate
GUT symmetries $SO(10)$ and $SU(5)$ that is a relevant scheme for
producing cosmic strings, which are also of cosmological interest,
because of the breaking of the accompanying abelian
groups~\cite{Vilenkin}. Obviously to unify couplings of SM
according to $SU(5)$ also contributions of supersymmetry (SUSY)
are needed~\cite{Altarelli}. Therefore the succession of
symmetries that we consider is
\begin{eqnarray}
E_6{\rightarrow}SO(10)\times\overline{U}(1){\rightarrow}SO(10)
{\rightarrow}SU(5)\times\widetilde{U}(1){\rightarrow}\nonumber \\
SU(5){\rightarrow}{SU(3)_C}\times{SU(2)_L}\times{U(1)}
{\rightarrow}{SU(3)_C}\times{U(1)_{em}}\
\end{eqnarray}

The fermions in the fundamental 27-plet are
distributed~\cite{Slansky} according to the representations of
$SO(10)\times\overline{U}(1)$
\begin{equation}
27=16^{1/4}+10^{-1/2}+1^1
\end{equation}
and then to those of $SU(5)\times\widetilde{U}(1)$ through
\begin{eqnarray}
16=\overline{5}^{3/2}+10^{-1/2}+1^{-5/2}\nonumber \\
10=5^1+{\overline{5}}\hfill^{-1}\nonumber \\ 1=1^0
\end{eqnarray}

The gauge bosons are in the self-adjoint representation 78 which
decomposes in $SO(10)\times\overline{U}(1)$ as
\begin{equation}
78=45^0+1^0+16^{-3/4}+\overline{16}^{3/4}
\end{equation}
with the subsequent $SU(5)\times\widetilde{U}(1)$ components
\begin{eqnarray}
45=24^0+10^2+{\overline{10}}\hfill^{-2}+1^0\nonumber \\
\overline{16}=1^{5/2}+\overline{10}^{1/2}+5^{-3/2}
\end{eqnarray}

The Higgs fields responsible for the breakings shown in Eq.(1) may
be in the representations 27 and 78, but it is necessary at least
one more to give masses to all the fermions through Yukawa terms
according to
\begin{equation}
27\times27=\overline{27}+351+351'
\end{equation}
which, for the purposes to be discussed in the next Sections, is
the 351 with $SO(10)\times\overline{U}(1)$ components
\begin{equation}
351=144^{1/4}+126^{-1/2}+54^1+16^{-5/4}+10^{-1/2}+1^{-2}
\end{equation}
that in terms of $SU(5)\times\widetilde{U}(1)$ are
\begin{eqnarray}
144=45^{-3/2}+40^{1/2}+24^{5/2}+\overline{15}^{1/2}+
\overline{10}^{1/2}+5^{-3/2}+{\overline{5}}\hfill^{-7/2}\nonumber
\\
126=50^{-1}+45^1+\overline{15}^3+10^{-3}+{\overline{5}}\hfill^{-1}+1^{-5}\nonumber
\\ 54=24^0+15^2+{\overline{15}}\hfill^{-2}
\end{eqnarray}
\section{Alternative useful for muon anomalous magnetic moment}

For the six breakings of Eq.(1) we use eight expectation values of
Higgs fields, which for economy will be taken in the
representations 78 and 351, to give masses to all fermions and
mixing of ordinary with exotic ones.

78 has no influence on fermions as is seen in Eq.(6), but is
needed to break $E_6$ because it contains $1^0$ which is invariant
under $SO(10)\times\overline{U}(1)$ and to break $SO(10)$ through
$45^0$ which contains $1^0$ of $SU(5)\times\widetilde{U}(1)$.

We use all the $SO(10)\times\overline{U}(1)$ representations of
351. $1^{-2}$ is necessary to break $SO(10)\times\overline{U}(1)$
giving mass to $L$ which will be therefore lighter than $X$ gauge
bosons, with the possible exception of the $\overline{Z}$
associated to $\overline{U}(1)$, but heavier than the other 10
exotic fermions. $126^{-1/2}$ is used to break
$SU(5)\times\widetilde{U}(1)$ giving mass to $\nu^c$, noting that
one has to take for the Higgs the complex conjugate of its
$SU(5)\times\widetilde{U}(1)$ representations according to Eq.(6).

To complete the breakings at GUT scale of $SU(5)$ we use $54^1$
and $144^{1/4}$ which give mass to exotic fermions and mix them
with ordinary ones, respectively. This is because they both
contain, according to Eq.(8), a 24 of $SU(5)$ which has a
component invariant under ${SU(3)_C}\times{SU(2)_L}\times{U(1)}$
that gives~\cite{Barbieri} the Yukawa couplings
\begin{eqnarray}
\phi(54,24)(D^c D-\frac{3}{2}E^c E-\frac{3}{2}N^c N)\nonumber \\
\phi(144,24)(d^c D-\frac{3}{2}E^c e-\frac{3}{2}N^c\nu)
\end{eqnarray}

Due to the fact that mass and mixing terms are analogous, which
would not happen in case of including $351'$, at GUT scale the
massless states apart from {\it u} will be
\begin{eqnarray}
d_0=d, \hspace{3mm}
{{d^c}_0}={d^c}\cos{\theta}+{D^c}\sin{\theta}\nonumber
\\ {e_0}^c=e^c, \hspace{3mm} { e_0}=e\cos{\theta}+E\sin{\theta}\nonumber \\
\nu_0=\nu\cos{\theta}+N\sin{\theta}
\end{eqnarray}
with orthogonal heavy mass states $\hat{E}$ etc.

The peculiar feature of equal mixing, even with large $\theta$, of
$e$ and $\nu$ and no mixing of $d$ and $u$ leads to unchanged
charged weak interactions for ordinary fermions, and additionally
Eq.(10) produces no change~\cite{Rizzo} in the neutral current
ones.

At this stage we may already anticipate that this scheme will not
allow $L$ to be source of UHECR. In fact, if the mass of
$\overline{Z}$ is of the same order but lower than that of $L$ the
latter will decay according to Fig.1
provided there is mixing of L and $\nu^c$ because of Eqs.(2, 4).
With the approximation $M_{\nu^c}<<M_L$, the lifetime $\tau_L$
would be
\begin{equation}
{\tau_L}^{-1}\simeq\frac{k}{4} M_L
\frac{1}{\lambda}(1-3{\lambda}^2+2{\lambda}^3), \hspace{3mm}
{{\lambda}}=\frac{M_{\overline{Z}}^2}{M_{L}^2}\hspace{2mm},
\end{equation}
where {\it k} will depend on the mixing. In case that this is
large $k/4\sim10^{-2}$ because it contains a coupling
${\alpha}_{GUT}$ and with $M_L\sim10^{16}$ GeV the result would be
$\tau_L\sim10^{-38} sec!$

But even without mixing of $L$ and $\nu^c$ or if
$M_{\overline{Z}}>M_L$, the decay of $L$ would be fast enough
through Fig.2
which would not require mixing of $L$ and $\cal{E}$ with ordinary
fermions $\it{O}$ if gauge boson $X$ belongs to 16 according to
Eqs.(2, 4). But mixing ${\it{O}}$ and ${\cal{E}}$ is necessary for
the subsequent decay of the latter. With
$\lambda=M_{X}^2/M_{L}^2$, $G=\Gamma_{X}^2/M_L^2$ where
$\Gamma_{X}$ is the $X$ width and in the approximation
$M_L>>M_{\cal{E}}, m_{\it{O}}$ the lifetime is
\begin{equation}
\tau_{L}^{-1}\simeq\frac{\alpha_{GUT}^2}{24\pi}\frac{M_{L}^5}{M_{X}^4}
\lambda^2 \int_{0}^{\frac{1}{2}}dx  x^2
\frac{3-4x}{(1-2x-\lambda)^2+\lambda G}\hspace{3mm},
\end{equation}
which, taking $M_X\sim10^{17}$ GeV, gives $\tau_L\sim10^{-32}$sec.

Going to the breaking of SM, this may be done by a Higgs in
$10^{-1/2}$ giving mass to ordinary fermions and mixing $L$ with
$N$ according to
\begin{equation}
H(10,\overline{5})(d^c d +e^c e+N^c L)+H(10,5)(u^c u+\nu^c \nu+LN)
\end{equation}
The mixing $L-N$ allows another possible channel of decay of the
former but less important than those of Fig.1, in case of large
$\it{k}$, and Fig.2 because it necessarily occurs at EW scale.

Moreover the $16^{-5/4}$ through its $SU(5)\times\widetilde{U}(1)$
component $1^{-5/2}$ gives a mixing of $L$ with $\nu^c$ which is
small if it occurs at this EW scale, but it might be large if it
was produced at $SU(5)\times\widetilde{U}(1)$ breaking where such
expectation value was possible. However $16^{-5/4}$ is the only
component of 351 which is not necessary for this alternative.

The eventual contribution of new particles~\cite{Grifols} to the
MAM has renewed its interest due to the $2.6-\sigma$ discrepancy
between experimental and SM calculation
\begin{equation}
\Delta a_{\mu}= \Delta{(\frac{g-2}{2})}_{\mu}\simeq4\times10^{-9}.
\end{equation}

Recent explanations come from SUSY~\cite{Ellis} and perhaps extra
dimensions~\cite{Park} but also from the exotic
fermion~\cite{Kephart} of $E_6$.

In fact with the present scheme of breakings based on 351, the
Yukawa couplings for $e$ and $E$ of Eqs.(9, 13) through the mixing
of Eq.(10) will give a "flavour" changing coupling of the physical
four-component $e_0$ and $\hat{E}$ with the Higgs field $h$ added
to the vacuum expectation value
\begin{equation}
{{\cal{L}}_{FC}}=\kappa\overline{e}_{0}(\alpha-\beta\gamma_5)\hat{E}h+h.c.\hspace{3mm},
\end{equation}
where $\kappa$ depends on the mixing angle and on the parameters
necessary to give the value of the lepton masses. An analogous
term for the second generation of fermions will produce a
contribution~\cite{Leveille} to MAM according to Fig.3

\begin{equation}
\Delta
a_{\mu}\simeq\frac{1}{8\pi^2}\frac{m_{\mu}}{M_M}\kappa^2F(\frac{M_h}{M_M})\hspace{3mm},
\end{equation}
where, for $M_h<<M_M$, $F(0)=\frac{1}{2}$. Therefore, being
${\kappa}\stackrel{<}{\sim}1$, to fill the present discrepancy
Eq.(14) it is necessary that the mass of the exotic lepton
$M_{M}\stackrel{<}{\sim }10^5$GeV. Since in the present scheme the
mixing is large because it occurs at GUT scale, it is not
unreasonable to have the above maximum values of $\kappa$ and
$M_M$.

It is interesting that in a different approach in which light
leptons are compled to heavy leptons and pions~\cite{Leite} a
contribution to $\Delta a_{\mu}$ analogous to Eq.(16) may be
consistent to UHECR due to the increase of $\nu$-nucleon
cross-section~\cite{Barshay}.

We may remark that the law of Eq.(16) evades the treatment of Ref.
~\cite{Einhorn} because our coupling Eq.(15) requires mixing of
two Higgs. On the contrary, the effective couplings coming from
SUSY or extra dimensions would give $\Delta
a_{\mu}\sim\frac{1}{16{\pi}^2} (\frac{m_{\mu}}{\Lambda})^2$ which,
to satisfy the experimental Eq.(14), would require a scale for new
phisics $\Lambda<1$ TeV. Therefore, corrections of MAM like
Eq.(16) are not pressed by a too close new physics scale, though
for the alternative explanations different masses of SUSY partners
or sum of Kaluza-Klein modes in extra dimension make their
situation easier.

\section{Alternative useful for ultra-high energy cosmic rays}

We choose now to break the symmetries of Eq.(1) using eight
expectation values of Higgs in all the
$SO(10)\times\overline{U}(1)$ components of 27 plus the minimum
needed ones of 78 and 351.

Regarding 78, we use as before its components $1^0$ to break $E_6$
and $45^0$, with $1^0$ of $SU(5)\times\widetilde{U}(1)$, to break
$SO(10)$. In addition we use again $45^0$, but with its $24^0$ of
$SU(5)\times\widetilde{U}(1)$, to break $SU(5)$ keeping the SM
symmetry and without influence on fermions.

Also two components of 351 are necessary: $1^{-2}$ to give mass to
$L$ and $126^{-1/2}$ for $\nu^c$ as before.

Now considering 27, $1^1$ is required to give mass to the 10
exotic fermions according to
\begin{equation}
\phi(1,1)(D^c D+E^c E+N^c N)
\end{equation}

Since presumably the expectation value of $1^1$ will appear at the
same scale of $1^{-2}$ corresponding to the breaking of
$SO(10)\times\overline{U}(1)$, it will depend on the constants of
the Yukawa couplings if $L$ is heavier or lighter than the exotic
fermions. It would be also possible that one of these expectation
values develops at a lower scale.

To break SM we need $10^{-1/2}$ which will give mass to ordinary
fermions and mixing of $L$ and $N$ as in Eq.(13).

Finally $16^{1/4}$ mixes ordinary and exotic fermions, which is
necessary because otherwise the latter would be stable, according
to
\begin{equation}
H(16,1)(d^c D+E^c e+N^c \nu)+H(16,\overline{5})(D^c d+e^c E+\nu^c
N)
\end{equation}

The expectation value of $SU(5)$ $\overline{5}$ may appear
breaking SM at the EW scale.

On its hand the $1^{-5/2}$ component of the first term of Eq.(18)
might appear earlier, at the breaking of
$SU(5)\times\widetilde{U}(1)$. If this occurs together with
Eq.(17) it would produce a mixing of the same type of Eq.(10) but
with smaller angle because of the higher scale of the mass term.
Therefore a discussion analogous to that of Sect.3 will lead to a
contribution to MAM smaller than the experimental discrepancy
Eq.(14) due to smaller mixing and larger $M_M$ in Eq.(16).

Alternatively, if $H(16,1)$ appears at the same EW scale of
$H(16,\overline{5})$ the mixing of exotic and ordinary fermions
will be even smaller giving way to negligible corrections of
charged and neutral weak interactions and of MAM.

But now the situation regarding UHECR will be different because
$L$ cannot decay to $\nu^{c}$ since there is no mixing between
them, and if $M_L<M_{\cal{E}}$, the decay of $\it{L}$ will be
given by $L\rightarrow \it{O}\it{O}\overline{\it{O}}V$ with $V$
vector boson of $SO(10)$ 45 which includes the twelve of SM as is
seen in Fig.4.
The coupling $\cal{E}{\it {OV}}$ is possible, because of
$\overline{U}(1)$ charge conservation of Eqs.(2, 4), only if there
is mixing of $\cal{E}$ and $\it{O}$.

If e.g. the scale of $27(1^1)$ is of the order of breaking of
$SO(10)\times\overline{U}(1)$ but that of $351(1^{-2})$ is
smaller, it is possible to have $M_L\sim10^{12}$ GeV corresponding
to UHECR whereas $M_{\cal{E}}\sim10^{16}$GeV and
$M_X\sim10^{17}$GeV due to breaking of $E_6$. Therefore the decay
of Fig.4 could be replaced by an effective coupling
$\sim\frac{1}{{M_X}^2 M_{\cal{E}}} \overline{\it{O}} \gamma^{\mu}
L \overline{\it{O}} \gamma_{\mu} \gamma^{\nu} \it{O} V_{\nu}$.
Since $M_V<<M_L$, an estimation of $L$ lifetime is
\begin{equation}
{\tau_L}^{-1}\sim{\alpha_{GUT}}^2 \alpha_M 10^{-7}
\frac{{M_L}^8}{{M_X}^4 M_{\cal{E}}^2 M_{V}}\hspace{3mm}.
\end{equation}
Introducing the above masses and $M_V$ $\sim 10^{2}$ GeV Eq.(19)
would give
\begin{equation}
{{\tau_L}^{-1}}\sim\alpha_M\frac{10^7}{sec}\hspace{3mm},
\end{equation}
of the order of universe age $t_0\sim10^{18}$sec for
$\alpha_M\sim10^{-25}$, not unreasonable for the square of the
extremely small mixing between EW and GUT scales. Then $\it{L}$
might be origin of UHECR.
\section{Conclusions}
We have seen that one possible alternative for Higgs responsible
for the breaking of $E_6$ symmetry passing through $SO(10)$ and
$SU(5)$ gives a strong mixing of ordinary and exotic leptons which
might allow a not too high mass of the latter and consequently a
correction of the $\mu$ magnetic moment of the order of the
discrepancy between experimental measurement and Standard Model
calculation. The qualitative difference with other explanations
based on supersymmetry or extra dimensions is that in our case the
scale of new physics is well above the TeV region.

Another alternative of Higgs for the same chain of symmetry
breakings starting from $E_6$ produces very heavy fermions in the
$SO(10)$ decuplet with extremely small mixing with ordinary ones,
so that the $\it{L}$ particle in $SO(10)$ singlet with mass
$\sim10^{12}$ GeV may have a lifetime of the order of universe age
and explain the ultra high energy cosmic rays. It is clear that to
avoid present overclosure of universe $\it{L}$ particles must have
been produced non-thermally to be now a small fraction of dark
matter.

Our proposal does not require non-renormalizable interactions for
the decay of $\it{L}$ at variance from other hypothetical
superheavy particles like cryptons~\cite{Kuzmin}, protected by
half-integer electric charge and a hidden gauge invariance, or
those coming from string models in a sector with broken GUT like
unitons and singletons, protected by a discrete
symmetry~\cite{Coriano}.

\bibliographystyle{unsrt}

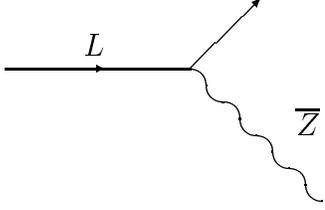
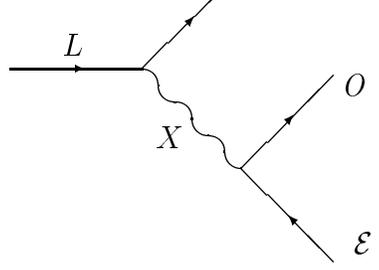
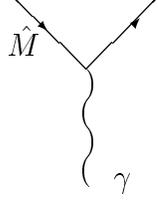
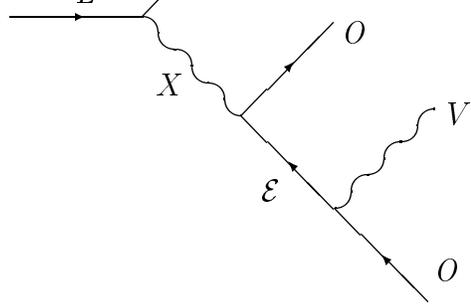
\end{document}